\documentclass[prb,twocolumn,groupedaddress,showpacs]{revtex4}

\usepackage[dvips]{graphicx}
\usepackage{color}
\usepackage{latexsym}
\usepackage{amsmath}
\usepackage{amssymb}
\usepackage{time}

\newcommand{\red}[1]{\textcolor{black}{#1}}

\begin{document}

\preprint{LA-UR-06-2457}

\title
   {Angle-resolved photoemission and first-principles electronic structure
   of single-crystalline $\alpha$-uranium (001)}

\author{C.P.~Opeil}
\affiliation{Department of Physics, Boston College,
   Chestnut Hill,
   MA 02467, USA}
\affiliation{Materials Science and Technology Division,
   Los Alamos National Laboratory,
   Los Alamos, NM 87545}

\author{R.K.~Schulze}
\affiliation{Materials Science and Technology Division,
   Los Alamos National Laboratory,
   Los Alamos, NM 87545}

\author{B.~Mihaila}
\affiliation{Materials Science and Technology Division,
   Los Alamos National Laboratory,
   Los Alamos, NM 87545}
\affiliation{Theoretical Division,
   Los Alamos National Laboratory,
   Los Alamos, NM 87545}

\author{H.M.~Volz}
\affiliation{Materials Science and Technology Division,
   Los Alamos National Laboratory,
   Los Alamos, NM 87545}

\author{J.C.~Lashley}
\affiliation{Materials Science and Technology Division,
   Los Alamos National Laboratory,
   Los Alamos, NM 87545}

\author{K.B.~Blagoev}
\affiliation{Theoretical Division,
   Los Alamos National Laboratory,
   Los Alamos, NM 87545}

\author{M.E.~Manley}
\affiliation{Materials Science and Technology Division,
   Los Alamos National Laboratory,
   Los Alamos, NM 87545}

\author{W.L.~Hults}
\affiliation{Materials Science and Technology Division,
   Los Alamos National Laboratory,
   Los Alamos, NM 87545}

\author{R.J.~Hanrahan Jr.}
\affiliation{Materials Science and Technology Division,
   Los Alamos National Laboratory,
   Los Alamos, NM 87545}

\author{R.C.~Albers}
\affiliation{Theoretical Division,
   Los Alamos National Laboratory,
   Los Alamos, NM 87545}

\author{P.B.~Littlewood}
\affiliation{Cavendish Laboratory,
   J.J. Thomson Avenue,
   Cambridge CB3 0HE,
   United Kingdom}

\author{J.L.~Smith}
\affiliation{Materials Science and Technology Division,
   Los Alamos National Laboratory,
   Los Alamos, NM 87545}


\begin{abstract}
Continuing the photoemission study begun with the work of Opeil
\emph{et al.} [Phys.\ Rev.\ B \textbf{73}, 165109 (2006)], in this
paper we report results of an angle-resolved photoemission
spectroscopy (ARPES) study performed on a high-quality
single-crystal $\alpha$-uranium at 173~K. The absence of
surface-reconstruction effects is verified using X-ray Laue and
low-energy electron diffraction (LEED) patterns. We compare the
ARPES intensity map with first-principles band structure
calculations using a generalized gradient approximation (GGA) and we
find good correlations with the calculated dispersion of the
electronic bands.
\end{abstract}

\pacs{
      79.60.-i,  
      71.20.Gj,  
      71.27.+a   
}

\maketitle

\section{Introduction}

To understand the effects of electronic correlations is one of the
principal challenges in the theory of metals~\cite{abrikosov,fulde},
and the actinide elements in the periodic table allow systematic
exploration of the role of electron correlation effects. With
increasing atomic number the f-orbitals shrink, so that Coulomb
interactions become increasingly dominant along the series beginning
with Th. The eventual dominance of the Coulomb repulsion over the
electronic kinetic energy produces a transition between the
itinerant metallic states of the early actinides and the
predominantly localized f-states of the later actinides, beyond Pu.
As well as a strong reorganization of the electronic structure near
the transition, a characteristic feature of this regime is strong
electron-lattice coupling as the different electronic configurations
cause complex crystal structures.

\begin{figure}[b]
   \includegraphics[width=3in]{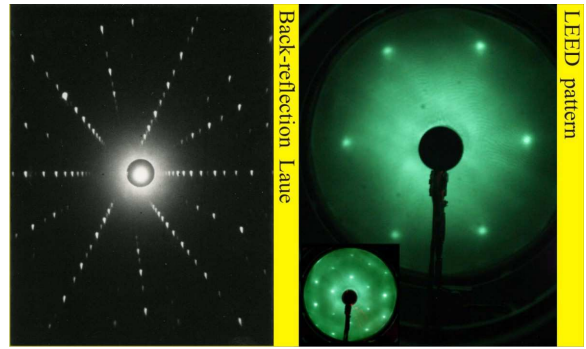}
   \caption{\label{fig1}
   LEED patterns at two energies (75 and 150~eV, respectively,
   the latter referring to the inset)
   and X-ray Laue diffraction pattern from the
   (001) $\alpha$-U single crystal at room temperature.
   The LEED and Laue patterns characterize the long-range atomic order of the
   crystal on two different spatial scales.
   }
\end{figure}

Uranium lies close to the boundary but with predominantly itinerant
character, and as the heaviest natural element is fundamental to the
study of nuclear and heavy-fermion physics~\cite{Lander_AP}. In the
$\alpha$ phase (i.e. the crystal structure of uranium below 935~K),
uranium undergoes a series of low-temperature instabilities that are
thought to arise from strong electron-phonon coupling present in
$\alpha$-U. This coupling is also responsible for other unusual
physical properties, such as the anisotropic thermal
expansion~\cite{Lloyd_66} and the strong temperature dependence of
the elastic moduli~\cite{Fisher}. Recently, it was reported that the
phonon spectrum of $\alpha$-U exhibits an unusually large thermal
softening of the phonon frequencies, suggesting that thermal effects
on the electronic structure in $\alpha$-U are more significant
thermodynamically than classical volume effects~\cite{manley_01}.
Furthermore, interest in the study of $\alpha$-U properties has been
stimulated by the advent of a 
generation of high-quality $\alpha$-U single crystals (see
Refs.~\cite{crystals,lashley_01} for details regarding the
preparation and purity of these crystals). Owing to their superior
properties, these $\alpha$-U single crystals have been the subject
of many recent
investigations~\cite{lashley_01,others1,others2,others3,others4,others5}.

\section{Angle-resolved photoemission of a (001) $\alpha$-uranium single crystal}

Photoemission spectroscopy has the ability to probe in detail the
electron-energy dispersions in the solid
and band structure mapping through angle-resolved photoemission
spectroscopy (ARPES) offers an ideal testbed for \emph{ab initio}
theoretical approaches to ground-state properties and electronic
correlations in metals. Past experiments on
U~\cite{photo1,photo2,photo3,photo4} have suffered from ambiguity
caused by measurements made on poorly defined or characterized
surfaces, caused by either contaminations issues or possible
formation of phases incompatible with any of the known bulk phases
of uranium metal (orthorombic $\alpha$, tetragonal $\beta$, and
$bcc$ $\gamma$ phases) in the case of U thin-film
formation~\cite{photo3}. Uncertainties related to the chemical
interaction between the overlayer and the substrate in thin-film
deposition studies, for example, make difficult the direct
comparison between theory and experiment. Using large $\alpha$-U
samples and a thorough sputter-anneal regimen we have overcome these
difficulties. In this paper, we report 
high-resolution ARPES data on high-quality $\alpha$-U
single-crystals at~173~K.

\begin{figure}[t]
   \includegraphics[width=3in]{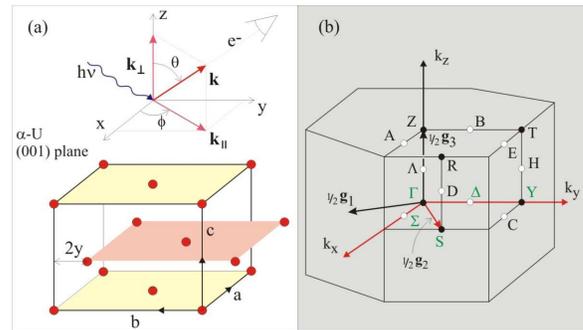}
   \caption{\label{fig2}
   (a) Schematic of the photoemission experiment relative to the
   single-crystal orientation~\cite{lattice}.
   (b) Brillouin zone corresponding to the $\alpha$-U single
   crystal \red{(see Fig.~3.6b in Ref.~\cite{space}).
   Here, $\mathbf{g}_i$ indicate the reciprocal lattice vectors,
   $\mathbf{g}_1 = 2\pi/(ba) (b,-a,0),\, \mathbf{g}_2 = 2\pi/(ba) (b,a,0),\,
   \mathbf{g}_3 = 2\pi/c (0,0,1)$.}
}
\end{figure}

Ultraviolet photoelectron spectra were recorded with a resolution of
28.5~meV using a Perkin-Elmer/Physical Electronics Model 5600 ESCA
system equipped with a monochromated Al~\textit{K}$\alpha$
(1486.6~eV) XPS source, a SPECS UVS 300 ultraviolet lamp (HeI,
$h\nu$=21.2~eV), and a spherical capacitor analyzer. The vacuum
chamber, which had a base pressure of $10^{-12}$~torr, was equipped
with a variable temperature sample stage of the range 150--1273~K.
Prior to the ARPES experiments, the surface preparation consisted of
repeated cycles of Ar ion sputtering and annealing at 873~K. As
temperature is reduced to 673~K, the surface reorders, and a
low-energy electron diffraction (LEED) pattern appears, see
Fig~\ref{fig1}. Major contaminant indicators, oxygen (O1s) and
carbon (C1s) signals in the XPS spectra were below the detection
limit ($<$ 1 at. \%)~\cite{prb_paper}. Surface cleanliness was
carefully examined to insure that all features in the ARPES are due
to the intrinsic $\alpha$-U surface and not surface contamination.
To assure that the ARPES results reflect the bulk properties of
$\alpha$-U, we used X-ray diffraction and LEED to study the
structure of the surface at room temperature. As shown in
Fig.~\ref{fig1}, our data found no detectable structural distortions
and show that the $c$ axis is perpendicular to the platelet surface
(see Brillouin zone depicted in Fig.~\ref{fig2}).

The crystal surface is aligned perpendicular to the analyzer, which
is set at an acceptance angle of $\pm$2 degrees to optimize the
instrumental sensitivity. We choose a particular azimuthal angle
$\phi$ in order to specify the direction to be probed in the (001)
plane. We also vary the polar angle $\theta$ to specify the
components of the \emph{in-plane} momentum component, $k_\|$, as
illustrated in Fig.~\ref{fig2}a. The polar angle was varied in steps
of 1$^\circ$ between 0$^\circ$ (normal emission) and 30$^\circ$, and
in steps of 5$^\circ$ between 30$^\circ$ and 60$^\circ$, with an
estimated angular error of $\pm$0.5$^\circ$. The energy-distribution
curve has been measured at each electron emission angle. Three sets
of data were obtained for the azimuthal angles corresponding to
$\tilde \Gamma\, \tilde \Sigma$, $\tilde \Gamma\, \tilde \Delta$,
and $\tilde \Gamma\, \tilde S$ directions~\cite{notation} in the
(001) plane (see Fig.~\ref{fig2}b~\cite{space}).

In contrast with our previous photoemission study in which we have
investigated only photoemission ($\theta=0$) along the $\tilde
\Gamma$ direction~\cite{prb_paper}, in this work we have not
attempted to carry out ARPES measurements for both HeI and HeII
energy excitations. It is true that, because of cross section
effects, the atomic photoionization energies at HeI and HeII
energies are such that the HeII spectra are more sensitive to
$f$-electron physics, whereas any $d$-electron feature will be
enhanced in the HeI spectra. However, the electronic mean-free path
is probably close to its minimum value for the HeII spectra, and
hence any surface state will be enhanced relative to bulk states for
the HeII spectra. Therefore, we have opted for performing only HeI
measurements. This is also the photoionization energy for which the
resolution of our experimental setup is optimal.

\section{First-principles electronic structure
calculations using the GGA/FLAPW method}

The photoemission data are compared with results of first-principles
band-structure calculations performed using the generalized gradient
approximation approach (GGA)~\cite{GGA} in the full-potential
linearized-augmented-plane-wave (FLAPW) method~\cite{Blaha}, with
added local orbitals for a better variational flexibility in the
radial basis functions~\cite{kunes}. The core states
are treated fully relativistically, whereas the valence $d$ and $f$
states relativistic effects are implemented using a second-order
variational method including spin-orbit coupling~\cite{so}, and
using the scalar-relativistic orbitals as basis. Our band structure
results are consistent with previous first-principles calculations
of properties of uranium metal via the full potential version of the
linear muffin-tin orbital (FP-LMTO) method~\cite{Umetal}. The
GGA/FLAPW approach was shown recently to compare favorably with the
X-ray photoemission spectroscopic (XPS) measurements on the same
single crystal of $\alpha$-U at room temperature~\cite{prb_paper}
and reproduces the spin-orbit splitting of the 6p core levels in
$\alpha$-U. The ground-state structure for $\alpha$-uranium is the
orthorhombic space group \emph{Cmcm} (no. 63), with uranium atoms
located at the 4c positions: (0, y, $\frac{1}{4}$) and (0, -y,
$\frac{3}{4}$) plus C-centering. In our calculations, we have used
the experimental lattice parameters $a$=2.858~$\rm{\AA}$,
$b$=5.876~$\rm{\AA}$, $c$=4.955~$\rm{\AA}$, and
$y$=0.105~\cite{lattice}. This structure is shown in
Fig.~\ref{fig2}a, where for clarity we have translated the origin of
the unit cell to coincide with an atom position.

\begin{figure}[t]
   \includegraphics[width=3in]{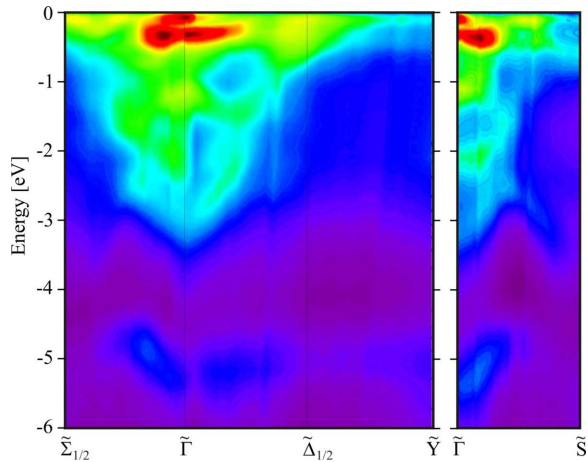}
   \caption{\label{fig3}
   Intensity map of the ARPES spectrum from  the (001) $\alpha$-U
   single crystal.
   Here, the color scale runs between violet (low intensity) and
   red (high intensity).
   \red{
   The symbols $\tilde \Sigma_\alpha$ and $\tilde \Delta_\alpha$ indicate the symmetry
   lines $(\alpha,\alpha,0)$ and  $(-\alpha,\alpha,0)$ depicted
   in Fig.~\ref{fig2}. See also Ref.~\cite{notation}.}
}
\end{figure}

\begin{figure}[t]
   \includegraphics[width=3in]{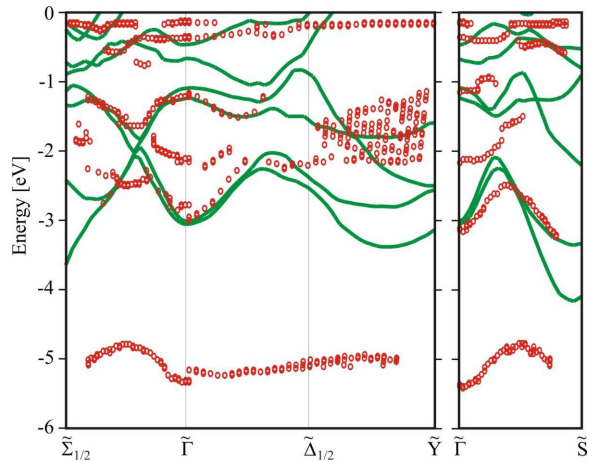}
   \caption{\label{fig4}
   Positions of the local maxima in the ARPES intensity map,
   together with the corresponding calculated band structure.
   \red{Assuming a free-electron final-state model, the value of
   $k_\perp$ was obtained using a zero inner-potential value, $V_0$
   (see the discussion in Ref.~\cite{notation}for further details).}
}
\end{figure}

\begin{figure}[t]
   \includegraphics[width=3in]{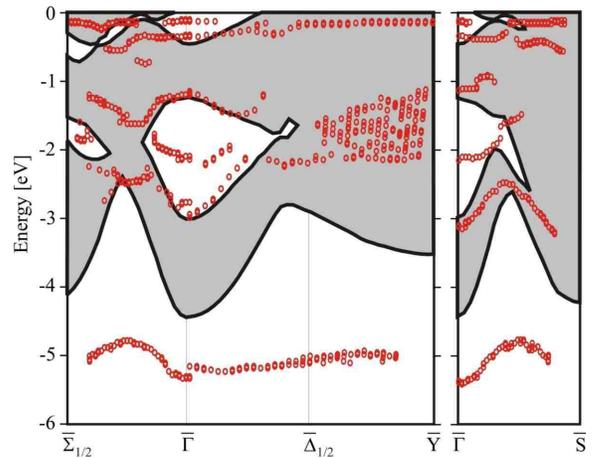}
   \caption{\label{fig5}
   Positions of the local maxima in the ARPES intensity map,
   together with the \red{projection of the $\alpha$-U bulk-derived bands
   onto the (001) surface Brillouin zone.
   The shaded regions indicate the range of values where $\alpha$-U
   energy bands exist when projected onto the $\bar \Gamma$ to $\bar \Sigma$,
   $\bar \Gamma$ to $\bar \Delta$, and $\bar \Gamma$ to $\bar S$ directions.}
   The white areas indicate the gaps in the bulk
   electronic band structure in which a surface state might exist.}
\end{figure}

\section{Results and Discussions}

In an ARPES experiment, the binding energy and crystal momentum of
the electron in the solid are related to the frequency of the
incident photon via the total energy and momentum conservation laws,
i.e.
\begin{align}
   E_\mathbf{KE} = & \
   h \nu - \Phi - |E_\mathbf{b}|
   \>,
   \\
   k_\| = & \
   \sqrt{(2m/h^2) \ E_\mathbf{KE}} \ \sin \theta
   \>,
\end{align}
where $\Phi$ is the \emph{work function} of the
spectrometer. These two equations, valid in the
noninteracting-electron approximation, constrain only the electron
kinetic energy $E_\mathbf{KE}$ and the \emph{in-plane} component of
the electron momentum, $\mathbf{k}_\|$, whereas the value of the
electron momentum perpendicular to the sample surface, $k_\perp$, is
not well defined because of the termination of the translational
symmetry normal to the sample surface. The uncertainty of $k_\perp$
is expected to be $\sim 0.2\div0.3$~\AA$^{-1}$ for a photon energy
$h\nu \sim 20$~eV on the basis of the mean-free path of electron in
solids, $\sim~3\div5$~\AA, which is significantly smaller than the
wave-vector along the c axis of $\alpha$-U, $2\pi / c
\sim$~1.2~\AA$^{-1}$, and can be neglected. Then, the intensity
measured in the ARPES experiment, $I(\mathbf{k}_\|,\omega)$, is
proportional to the electronic density of states, weighed by the
square of the transition matrix element, $|M_{fi}^\mathbf{k}|^2$, of
the photon-electron interaction between the initial and final states
of the N-particle system, i.e.
\begin{align}
   I(\mathbf{k}_\|,\omega)
   \propto & \
   |M_{fi}^\mathbf{k}(\nu,\hat \epsilon)|^2 \
   \mathrm{N}(\mathbf{k},\omega) \
   f(\omega)
   \>,
\label{intensity}
\end{align}
where $|M_{fi}^\mathbf{k}(\nu,\hat \epsilon)|^2$ depends on the
electron momentum and on the energy ($h \nu$) and polarization
($\hat \epsilon$) of the incident photon.
$\mathrm{N}(\mathbf{k},\omega)$ is the electron
directional-dependent density of states of the electron, and the
Fermi function, $f(\omega) = (e^{\hbar \omega/k_B T}+1)^{-1}$, is
included in Eq.~\eqref{intensity} because photoemission spectroscopy
probes only the occupied electronic states. While a quantitative
analysis of the ARPES experimental results, requiring the detailed
modeling of the structure function, is beyond the scope of this
study, we find interesting correlations between the experimental
intensity map, $I(\mathbf{k}_\|,\omega)$, and bulk electronic band
structure calculated using the GGA/FLAPW method.

Figure~\ref{fig3} shows the $\alpha$-U single-crystal ARPES data at
HeI photon energy as a function of the component of the electron
momentum in the (001) plane, $k_\|$. For comparison, in
Fig.~\ref{fig4} we plot the positions of the intensity maxima in the
ARPES intensity map together with the corresponding electronic band
structure. We observe good correlations between the ARPES and the
calculated band structure, which tracks the main features of the
photoemission landscape in the (001) plane. However, the
high-intensity spectral structures located close to the Fermi
surface are shifted with respect to the high-density level regions
in the band structure. We also notice several features which do not
have a correspondent in the calculated bulk electronic band
structure. In particular, we confirm the presence of peaks located
at -0.1~eV, -2.2~eV along the $\tilde \Gamma$ direction (see also
Ref.~\cite{prb_paper}).

For a photon at the HeI energy, $h\nu$=21.2~eV, the plot of the
``universal'' escape depth \red{(see chapter~1 in
Ref.~\cite{hufner})} indicates the escape depth in our experiment is
less than 10~\AA, which must be compared with the lattice spacing in
the $c$~direction ($c$=4.955~$\rm{\AA}$). Therefore, in an ARPES
experiment at this photon energy, we probe at most 2-3 atomic layers
perpendicular to the (001) surface, and surface scattering may
contribute to the measured photoemission spectrum. While the dirt
contributions due to surface states have been minimized through
careful surface preparation~\cite{prb_paper}, we may still observe
the presence of surface states located in the gaps of the bulk band
structure~\red{(see chapter~8 in Ref.~\cite{hufner} and
Refs.~\cite{tamm,shockley})}.
The shaded region shown in Fig.~\ref{fig5} indicates the range of
values where $\alpha$-U energy bands exist when projected onto the
(001) plane. The white areas indicate possible regions where surface
states might exist.

The GGA/FLAPW approximation used here, captures part of the
electronic correlations, but is not able to describe the
strongly-varying electronic density in heavy-fermion
materials~\cite{fulde}. Therefore, the discrepancies between our
calculations and the observed ARPES spectra can be attributed in
part to the strong correlations in the electronic liquid or yet
unidentified contribution of surface or collective states.
A recent analysis of the specific-heat data below 10~K, measured on
$\alpha$-U samples of the same pedigree~\cite{lashley_01}, gives an
electronic specific heat
$\gamma_\mathrm{exp}$=9.13~mJ~K$^{-2}$~mol$^{-1}$ and a
low-temperature limiting Debye temperature $\Theta_D$=256~K. Using
these values, and taking the band structure
$\gamma_\mathrm{b.s.}$=5.89~mJ~K$^{-2}$~mol$^{-1}$, we obtain an
upper bound on the mass-enhancement factor~\red{\cite{eph}},
$m^*/m_\mathrm{b.s.} = \gamma_\mathrm{exp}/\gamma_\mathrm{b.s.} =
1.55$, for $\alpha$-U single crystal. This modest fermi edge mass
enhancement could of course have its origin in either or both of the
electron-phonon coupling or electron correlation effects, which go
beyond those included in the GGA/FLAPW band-structure calculations.
If the former, the redistribution of spectral weight will disappear
rapidly off-shell, whereas the electron-electron interactions would
play an increasing role at higher energies. The generally good
correlations between experiment and theory suggest that correlations
effects are not very large in $\alpha$-U, and that it is appropriate
to think of it as a band metal, and not a highly-correlated system.

%

\section{Concluding remarks}

In summary, in this paper we report 
high-resolution ARPES measurements of a high-quality (001)
$\alpha$-U single crystal and compare them with first-principles
electronic band structure calculations using the GGA/FLAPW method.
Together with detailed theoretical studies of electronic correlation
effects, further photoemission studies at a synchrotron radiation
source (necessary to take advantage of an improved energy resolution
and the ability to modify the incident photon energy) are being
pursued in order to understand quantitatively the band structure
using ARPES in $\alpha$-U single crystals.


\begin{acknowledgments}
This work was supported in part by the LDRD program at Los Alamos
National Laboratory. B.M. acknowledges financial support from ICAM.
The authors gratefully acknowledge the contribution of the U(001)
samples from: H.F.~McFarlane, K.M.~Goff, F.S.~Felicione,
C.C.~Dwight, D.B.~Barber, C.C.~McPheeters, E.C.~Gay, E.J.~Karell and
J.P.~Ackerman at Argonne National Laboratory. B.M. would like to
thank D.L. Smith, J.M. Wills, M.D. Jones, and I. Schnell for useful
discussions.
\end{acknowledgments}


%
%
%
%
%

\end{document}